\documentclass[letterpaper,english,aps, prl]{revtex4-1}
\usepackage[T1]{fontenc}
\usepackage[latin9]{inputenc}
\usepackage{babel}
\usepackage{units}
\usepackage{bm}
\usepackage{amsmath}
\usepackage{amssymb}
\usepackage{graphicx}
\usepackage[unicode=true,pdfusetitle,
 bookmarks=true,bookmarksnumbered=false,bookmarksopen=false,
 breaklinks=false,pdfborder={0 0 1},backref=false,colorlinks=false]
 {hyperref}



\begin{document}
\title{Tutorial: Concepts and numerical techniques for modeling individual
phonon transmission at interfaces}
\author{Zhun-Yong Ong}
\email{ongzy@ihpc.a-star.edu.sg}

\affiliation{Institute of High Performance Computing, A{*}STAR, Singapore 138632,
Singapore}
\begin{abstract}
At the nanoscale, thermal transport across the interface between two
lattice insulators can be described by the transmission of bulk phonons
and depends on the crystallographic structure of the interface and
the bulk crystal lattice. In this tutorial, we give an account of
how an extension of the Atomistic Green's Function (AGF) method based
on the concept of the Bloch matrix can be used to model the transmission
of individual phonon modes and allow us to determine the wavelength
and polarization dependence of the phonon transmission. Within this
framework, we can explicitly establish the relationship between the
phonon transmission coefficient and dispersion. Details of the numerical
methods used in the extended AGF method are provided. To illustrate
how the extended AGF method can be applied to yield insights into
individual phonon transmission, we study the (16,0)/(8,0) carbon nanotube
intramolecular junction. The method presented here sheds light on
the modal contribution to interfacial thermal transport between solids.
\end{abstract}
\keywords{Phonons, interfaces}
\maketitle

\section{Introduction}

Heat conduction between dissimilar non-metallic materials is at present
a topic of increasing relevance for technological development in nanoelectronics,
optoelectronics, nanomechanics and thermoelectrics because of the
heat dissipation issues associated with high temperatures and power
densities which limit device performance. One of the basic challenges
to efficient heat dissipation in semiconductors and insulators at
the nanoscale is the thermal boundary resistance which arises from
incomplete phonon transmission across the crystallographic interface.~\citep{EPop:NR10_Energy}
On the other hand, the impedance of thermal transport by interfaces
may be exploited for thermoelectric applications\ \citep{DGCahill:JAP03_Nanoscale,EPop:NR10_Energy,DGCahill:APR14_Nanoscale,PChen:PRL14}.
Therefore, a substantial amount of experimental and theoretical work~\citep{DGCahill:JAP03_Nanoscale,DGCahill:APR14_Nanoscale}
has been motivated by the desire to advance our understanding of interfacial
phonon transmission and its role in interfacial thermal transport. 

In most theories of phonon-mediated interfacial thermal transport,~\citep{GChen:Book05_Nanoscale}
it is assumed that energy is dissipated across an interface when an
incident \emph{bulk} phonon, propagating in one medium towards the
other, crosses the interface with the probability of transmission
given by the phonon transmission coefficient. This physical picture
underlies the two major acoustics-based analogies,~\citep{ETSwartz:RMP89_Thermal,GChen:Book05_Nanoscale}
the acoustic mismatch model (AMM) and the diffuse mismatch model (DMM),
commonly used to interpret experimental~\citep{RCostescu:PRB03_Thermal}
and simulation-based studies~\citep{PKSchelling:APL02_Phonon} of
interfacial thermal transport. However, in spite of their widespread
use in modeling thermal boundary resistance, they suffer from several
shortcomings. Firstly, they typically adopt an idealized model of
the phonon dispersion and ignore the contribution from optical phonon
modes. Secondly, they cannot determine the dependence of phonon transmission
on the atomistic structure of the interface as phonon scattering by
the interface is \emph{assumed} to be either completely specular (in
the AMM) or diffusive (in the DMM). 

On the other hand, the relationship between the atomistic structure
of the interface and phonon transmission can be studied using the
Atomistic Green's Function (AGF) method developed by Mingo and Yang,~\citep{NMingo:PRB03_Phonon}
a highly versatile lattice dynamics-based technique which can be coupled
to either empirical~\citep{WZhang:JHT07_Simulation,WZhang:NHT07}
or ab initio-based models of interatomic forces~\citep{ZTian:PRB12_Enhancing}
and has proved to be a powerful tool for studying ballistic phonon
transport in atomistic models;~\citep{WZhang:NHT07} the technique
has been applied to graphene grain boundaries~\citep{YLu:APL12_Thermal,AYSerov:APL13_Effect},
silicon-germanium heterostructures~\citep{WZhang:JHT07_Simulation,ZHuang:JHT11_Modeling,ZTian:PRB12_Enhancing}
and superlattices.~\citep{MNLuckyanova:Science12_Coherent} In addition,
significant recent progress has been made in the development of molecular
dynamics simulation techniques for determining the frequency-dependent
spectral content of the thermal boundary conductance.~\citep{KSaaskilahti:PRB14_Role,YZhou:PRB17_Full}
However, one of the principal drawbacks of the traditional AGF method~\citep{WZhang:NHT07,WZhang:JHT07_Simulation,ZHuang:JHT11_Modeling}
is its inability to describe individual phonon transmission explicitly
in terms of the bulk phonon dispersion, unlike the AMM where the individual
phonon transmission coefficients can be determined, potentially limiting
our understanding of the physical processes in nanoscale interfacial
thermal transport and their dependence on atomistic structure. 

However, a straightforward and \emph{efficient} extension of the existing
AGF methodology has been developed recently developed in Ref.~\citep{ZYOng:PRB15_Efficient}
to describe the ballistic transmission of individual phonon modes
and their dependence on polarization, frequency ($\omega$), and wave
vector ($\mathbf{k}$), connecting the transmission coefficient to
phonon dispersion. A key ingredient of this extension of the AGF method
is the notion of the \emph{Bloch matrix} which can be derived from
the \emph{surface} Green's function and yields the \emph{bulk} phonon
modes that constitute the individual transmission \emph{channels}.
Conceptually, this extension of the AGF method bridges the gap between
the existing AGF method, in which the connection between transmittance
and phonon dispersion is obscure, and the analytical theories of the
AMM and DMM, where the individual phonon transmission coefficients
can be computed but the dependence on the atomistic structure of the
interface is unclear. 

In this tutorial, we will not review the general phenomenon of nanoscale
interfacial thermal transport, as several excellent review articles
have been written,~\citep{ETSwartz:RMP89_Thermal,DGCahill:JAP03_Nanoscale,DGCahill:APR14_Nanoscale}
but rather, describe the technique of the aforementioned \emph{extension}
of the Atomistic Green's Function (AGF) method,~\citep{ZYOng:PRB15_Efficient}
which we will call the \emph{extended AGF method} in the rest of the
paper, for calculating the ballistic transmission of individual phonons.
It is hoped that after reading this tutorial article, the reader will
gain a better understanding of the practical steps involved in calculating
individual phonon transmission coefficients. The organization of the
rest of the tutorial is as follows. We first review and describe the
traditional AGF method in terms of the numerical inputs and how the
calculations are set up. Next, we describe how the Bloch matrices
are derived from the surface Green's functions used in the traditional
AGF method, and show how the Bloch matrices can be exploited to determine
the phonon eigenmode and velocity matrices which can be combined with
the Green's function of the scattering region to yield the transmission
matrices and individual phonon transmission coefficients. We demonstrate
the basic ideas of the extended AGF method using the simple first
example of a linear atomic chain junction. To illustrate the application
of the extended AGF method to more realistic systems, we simulate
phonon transmission across the (16,0)/(8,0) carbon nanotube intramolecular
junction~\citep{GWu:PRB07_Thermal} and analyze how the phonon transmission
coefficients depend on polarization, frequency and wavelength. 

\subsection{Review of traditional AGF method}

The AGF method is derived from the well-known nonequilibrium Green's
function (NEGF) method used in the modeling of tight-binding electron
transport~\citep{GKlimeck:APL95_Quantum,MBNardelli:PRB99_Electronic}
and many excellent pedagogical resources on the NEGF method have been
made available by S. Datta.~\citep{SDatta:Book97_Electronic,SDatta:SM00_Nanoscale,SDatta:Nanohub}
The key idea in the NEGF method~\citep{SDatta:Book97_Electronic}
is that in the Landauer-B{\"{u}}ttiker picture,~\citep{RLandauer:JPCM89_Conductance}
the electrical current in the channel arises from the transmission
of electrons between the leads and is determined from the energy-dependent
transmission function or transmittance computed from the Green's function
of the system. In actual numerical implementation, the Green's function
is evaluated from a tight-binding Hamiltonian model.~\citep{SDatta:SM00_Nanoscale} 

The application of the NEGF method to thermal transport in realistic
atomistic structures was first formulated by Mingo and Yang who used
the AGF method to describe phonon flow in an atomistic model of silica-coated
silicon nanowires.~\citep{NMingo:PRB03_Phonon} A more detailed and
highly readable account of the AGF approach is given in Ref.~\citep{NMingo:Springer09}.
Like in the electron NEGF method, the phonon current in the lattice
is determined by a frequency-dependent transmittance which depends
on the Green's function derived from the force-constant matrix $\mathbf{K}$
describing the vibrational character of the lattice, instead of the
tight-binding Hamiltonian. The nondiagonal elements of $\mathbf{K}$
are given by, for $i\neq j$,
\begin{equation}
K_{ij}=\frac{\partial^{2}E}{\partial u_{i}\partial u_{j}}\ ,\label{eq:ForceConstantMatrix}
\end{equation}
where $E$ is the total lattice energy and $u_{i}$ is the displacement
of the $i$-th atomic degree of freedom with respect to its equilibrium
value. The expression in Eq.~(\ref{eq:ForceConstantMatrix}) is just
the Hessian matrix and is symmetric. The acoustic sum rule implies
that the diagonal elements of $\mathbf{K}$ can be obtained from the
condition $K_{ii}=-\sum_{j}K_{ij}$.

If we take $\omega$ to be the vibrational frequency, then the equation
of motion in frequency space for the system is 
\begin{equation}
(\omega^{2}\mathbf{M}-\mathbf{K})\mathbf{u}=\mathbf{0}\label{eq:dynamicalEqn}
\end{equation}
where $\mathbf{M}$ is a diagonal matrix with its matrix elements
corresponding to the masses of the constituent atoms and $\mathbf{u}$
is a column vector with its elements corresponding to the individual
degrees of freedom $u_{i}$. Equation~(\ref{eq:dynamicalEqn}) can
be rewritten as an eigenvalue equation 
\begin{equation}
(\omega^{2}\mathbf{I}-\mathbf{H})\mathbf{\bar{u}}=\bar{\mathbf{0}}\label{eq:HarmonicEqnMotion}
\end{equation}
where $\mathbf{H}=\mathbf{M}^{-1/2}\mathbf{K}\mathbf{M}^{-1/2}$ is
the mass-normalized harmonic matrix and $\mathbf{\bar{u}}=\mathbf{M}^{1/2}\mathbf{u}$.
We can interpret Eq.~(\ref{eq:HarmonicEqnMotion}) as the frequency-space
equation of motion for the lattice, analogous to the Schr\"{o}dinger
equation which is the equation of motion for the electron wave function,
and regard $\mathbf{H}$ as the lattice-dynamical analog of the tight-binding
Hamiltonian. Typically, for a system with a finite number of degrees
of freedom, the eigenmodes, which are its stationary states, and eigenfrequencies
can be found by solving Eq.~(\ref{eq:HarmonicEqnMotion}). For an
infinite bulk system with translational symmetry, the eigenmodes are
called phonons and correspond to the periodic atomic displacements
in the lattice. However, the phonon transmittance is not determined
by simply solving Eq.~(\ref{eq:HarmonicEqnMotion}) which is an eigenvalue
equation. The problem of transmission is more complicated conceptually
and numerically as it deals with an infinitely large system that lacks
the translational symmetry like in a uniform bulk system. Rather,
treating phonon transmission involves determining how \emph{asymptotic}
bulk lattice wave excitations, bulk phonons in our case, pass through
a localized region and transit to asymptotic bulk lattice wave states
on the other side, and this calls on a different method of solution
such as the AGF method in which the primary object of study is the
\emph{transitions} between asymptotic bulk phonon states, which are
extended infinitely into the bulk, rather than the eigenstates of
the lattice. 

\begin{figure}
\includegraphics[width=8cm]{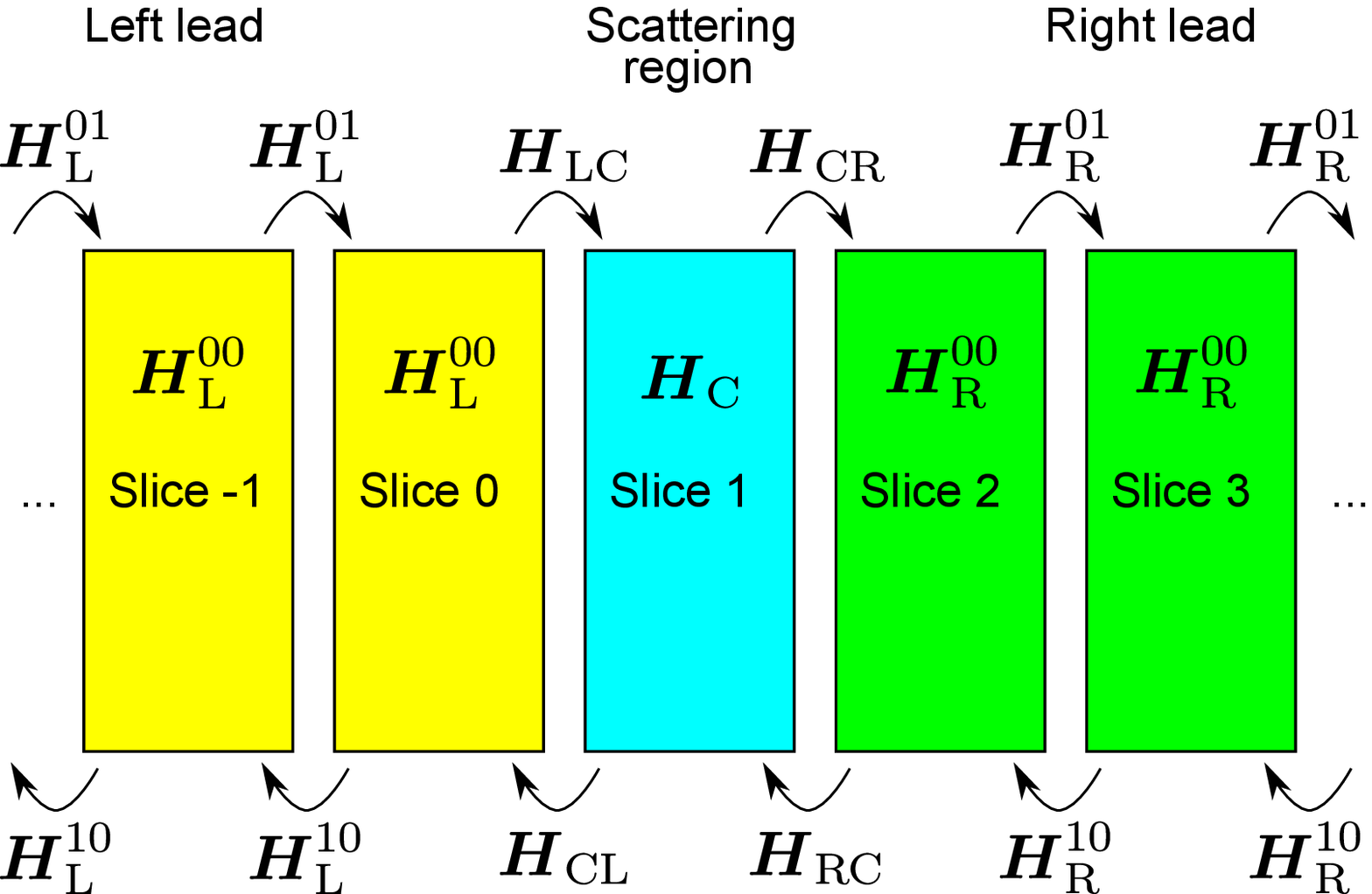}

\caption{Schematic of scattering system (left lead, scattering region and right
lead) and the submatrices associated with each slice or principal
layer which represents the set of atomic degrees of freedom for a
block row in Eq.~(\ref{eq:SystemForceConstantMatrix}). The left
and right lead each consist of a semi-infinite one-dimensional array
of identical slices while the scattering region corresponds to the
interface.}

\label{fig:SystemSchematic}
\end{figure}

\subsubsection{Arrangement of system into principal layers}

At the interface between two lattices, translational symmetry is broken
and the discontinuity in the crystallographic structure results in
phonon transmission and reflection by the interface. The AGF method
essentially computes the frequency-dependent transmittance for the
interface. In the AGF method, the harmonic matrix $\mathbf{H}$ is
partitioned into submatrices according to the physical arrangement
of the degrees of freedom within our simulation structure. In the
partition scheme shown in Fig.~\ref{fig:SystemSchematic}, there
are three subsystems: (1) the left lead, (2) the scattering region
and (3) the right lead. The leads correspond to the bulk lattices
while the scattering region contains their interface. The leads are
each arranged into a semi-infinite one-dimensional array of identical
slices (or principal layers) of equal size while the scattering region
is considered a slice by itself. The number of degrees of freedom
in each slice should be large enough so that only adjacent slices
can couple mechanically, and we characterize the spacing between the
slices in the lead by the lattice constant $a_{\alpha}$, where $\alpha=\text{L}$
and $\alpha=\text{R}$ for the left and right lead, respectively.
Thus, the entire system has an infinite number of slices, each of
which can be indexed by an integer that increases as one goes from
left to right slice-wise. In our convention, the scattering region
is defined as slice $1$ while the principal layers in the left and
right lead are enumerated $-\infty,\ldots,0$ and $2,\ldots,+\infty$,
respectively. 

Given our partitioning scheme, the harmonic matrix $\mathbf{H}$ from
Eq.~(\ref{eq:HarmonicEqnMotion}) can be structured as a block-tridiagonal
matrix, 
\begin{equation}
\mathbf{H}=\left(\begin{array}{ccccccc}
\ddots & \ddots\\
\ddots & \boldsymbol{H}_{\text{L}}^{00} & \boldsymbol{H}_{\text{L}}^{01}\\
 & \boldsymbol{H}_{\text{L}}^{10} & \boldsymbol{H}_{\text{L}}^{00} & \boldsymbol{H}_{\text{LC}}\\
 &  & \boldsymbol{H}_{\text{CL}} & \boldsymbol{H}_{\text{C}} & \boldsymbol{H}_{\text{CR}}\\
 &  &  & \boldsymbol{H}_{\text{RC}} & \boldsymbol{H}_{\text{R}}^{00} & \boldsymbol{H}_{\text{R}}^{01}\\
 &  &  &  & \boldsymbol{H}_{\text{R}}^{10} & \boldsymbol{H}_{\text{R}}^{00} & \ddots\\
 &  &  &  &  & \ddots & \ddots
\end{array}\right)\label{eq:SystemForceConstantMatrix}
\end{equation}
where $\boldsymbol{H}_{\text{C}}$, and $\boldsymbol{H}_{\text{CL}}$
($\boldsymbol{H}_{\text{CR}}$) are respectively the force-constant
submatrices corresponding to the interface region (slice 1) and the
coupling between the interface region and the semi-infinite left (right)
lead. We can associate each slice in Fig.~\ref{fig:SystemSchematic}
with a block row and column in $\mathbf{H}$. In the standard AGF
setup, the submatrices $\boldsymbol{H}_{\alpha}^{00}$ and $\boldsymbol{H}_{\alpha}^{01}$,
where $\alpha=\text{L}$ and $\alpha=\text{R}$ for the left and right
lead, respectively, characterize the lead phonons. In the each lead,
$\boldsymbol{H}_{\alpha}^{00}$ corresponds to the force-constant
submatrix for each slice while $\boldsymbol{H}_{\alpha}^{01}$ ($\boldsymbol{H}_{\alpha}^{10}$)
corresponds to the harmonic coupling between each slice and the slice
to its right (left) in the lead. In the rest of the paper, we reserve
$\alpha$ as the dummy variable for distinguishing the leads, with
$\alpha=\text{L}$ and $\alpha=\text{R}$ representing the left and
right lead, respectively. 

\subsubsection{Force-constant matrices and Green's functions }

We note here that in spite of the infinite number of slices making
up the system in Fig.~\ref{fig:SystemSchematic}, only a finite set
of unique force-constant matrices ($\boldsymbol{H}_{\text{C}}$, $\boldsymbol{H}_{\text{CL}}$,
$\boldsymbol{H}_{\text{CR}}$, $\boldsymbol{H}_{\text{L}}^{00}$,
$\boldsymbol{H}_{\text{L}}^{01}$, $\boldsymbol{H}_{\text{R}}^{00}$
and $\boldsymbol{H}_{\text{R}}^{01}$) are needed as inputs for the
AGF calculation because the leads are made up of identical slices
and the Hermicity of $\mathbf{H}$ implies that $\boldsymbol{H}_{\text{LC}}=(\boldsymbol{H}_{\text{CL}})^{\dagger}$,
$\boldsymbol{H}_{\text{RC}}=(\boldsymbol{H}_{\text{CR}})^{\dagger}$,
and $\boldsymbol{H}_{\alpha}^{01}=(\boldsymbol{H}_{\alpha}^{10})^{\dagger}$.
The periodic arraying of the slices in the leads means that each slice
constitutes a unit cell, but not necessarily the primitive unit cell,
and that the bulk phonon dispersion, which relates the vibrational
frequency $\omega$ to the phonon wave vector $k$, can be determined
from the eigenvalue equation 
\begin{equation}
[\omega^{2}\boldsymbol{I}_{\alpha}-\boldsymbol{D}_{\alpha}(k)]\boldsymbol{\phi}(k)=\boldsymbol{0}\ ,\label{eq:BulkPhononDispersion}
\end{equation}
where $\boldsymbol{D}_{\alpha}(k)=\boldsymbol{H}_{\alpha}^{10}e^{-ika_{\alpha}}+\boldsymbol{H}_{\alpha}^{00}+\boldsymbol{H}_{\alpha}^{01}e^{ika_{\alpha}}$
is the dynamical matrix and $\boldsymbol{I}_{\alpha}$ is the identity
matrix. 

\begin{figure}

\newcommand{\SyHC}{$\boldsymbol{H}_{\text{C}}^\prime$}
\newcommand{\SyHL}{$\boldsymbol{H}_{\text{L}}^\prime$}
\newcommand{\SyHR}{$\boldsymbol{H}_{\text{R}}^\prime$}
\newcommand{\SyHCL}{$\boldsymbol{H}_{\text{CL}}^\prime$}
\newcommand{\SyHCR}{$\boldsymbol{H}_{\text{CR}}^\prime$}
\newcommand{\SyHLC}{$\boldsymbol{H}_{\text{LC}}^\prime$}
\newcommand{\SyHRC}{$\boldsymbol{H}_{\text{RC}}^\prime$}

\newcommand{\SyUadvL}{$\boldsymbol{U}_\text{L}^\text{adv}(-)$}
\newcommand{\SyUretL}{$\boldsymbol{U}_\text{L}^\text{ret}(-)$}
\newcommand{\SyUadvR}{$\boldsymbol{U}_\text{R}^\text{adv}(+)$}
\newcommand{\SyUretR}{$\boldsymbol{U}_\text{R}^\text{ret}(+)$}

\newcommand{\SyVadvL}{$\boldsymbol{V}_\text{L}^\text{adv}(-)$}
\newcommand{\SyVretL}{$\boldsymbol{V}_\text{L}^\text{ret}(-)$}
\newcommand{\SyVadvR}{$\boldsymbol{V}_\text{R}^\text{adv}(+)$}
\newcommand{\SyVretR}{$\boldsymbol{V}_\text{R}^\text{ret}(+)$}

\includegraphics[width=8cm]{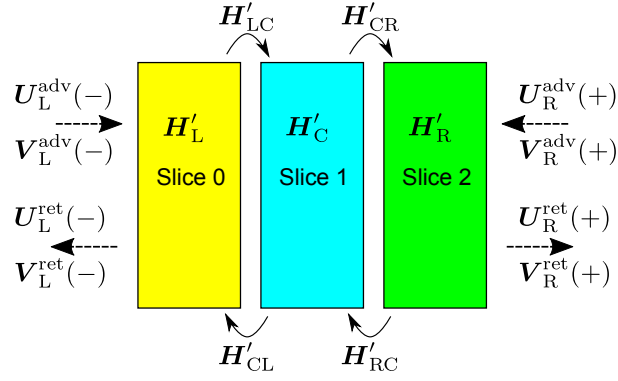}

\caption{Schematic of the finite projected system in Eq.~(\ref{eq:ProjectedForceConstantMatrix}),
consisting of the scattering region (slice $1$) and its terminated
edges (slices $0$ and $2$). The frequency-dependent dynamics of
the semi-infinite leads are implicitly included in $\boldsymbol{H}_{\text{L}}^{\prime}$
and $\boldsymbol{H}_{\text{R}}^{\prime}$ through the surface Green's
functions $\boldsymbol{g}_{\text{L},-}^{\text{ret}}$ and $\boldsymbol{g}_{\text{R},+}^{\text{ret}}$
from which we can derive the incoming and outgoing phonon modes {[}$\boldsymbol{U}_{\text{L}}^{\text{adv/ret}}(-)$
and $\boldsymbol{U}_{\text{R}}^{\text{adv/ret}}(+)${]} and their
group velocities {[}$\boldsymbol{V}_{\text{L}}^{\text{adv/ret}}(-)$
and $\boldsymbol{V}_{\text{R}}^{\text{adv/ret}}(+)${]}.}

\label{fig:ProjectedSystemSchematic}
\end{figure}

In principle, the system dynamics are determined by the infinitely
large $\mathbf{H}$ in Eq.~(\ref{eq:SystemForceConstantMatrix}).
However, if we restrict ourselves to describing the oscillatory motion
at frequency $\omega$, the problem becomes more tractable as we need
only to project the lattice dynamics onto a finite portion of the
system,~\citep{JSWang:EPJB08_Quantum,NMingo:Springer09} corresponding
to slices 0 to 2 in Fig.~\ref{fig:SystemSchematic}, to determine
the phonon transmittance through the scattering region (slice 1).
Hence, we can use the submatrices in Eq.~(\ref{eq:SystemForceConstantMatrix})
to construct the \emph{effective} frequency-dependent harmonic matrix
or `Hamiltonian' for this subsystem, as shown in Fig.~\ref{fig:ProjectedSystemSchematic}:~\citep{JSWang:EPJB08_Quantum}
\begin{equation}
\mathbf{H}^{\prime}=\left(\begin{array}{ccc}
\boldsymbol{H}_{\text{L}}^{\prime} & \boldsymbol{H}_{\text{LC}}^{\prime} & 0\\
\boldsymbol{H}_{\text{CL}}^{\prime} & \boldsymbol{H}_{\text{C}}^{\prime} & \boldsymbol{H}_{\text{CR}}^{\prime}\\
0 & \boldsymbol{H}_{\text{RC}}^{\prime} & \boldsymbol{H}_{\text{R}}^{\prime}
\end{array}\right)\ ,\label{eq:ProjectedForceConstantMatrix}
\end{equation}
where $\boldsymbol{H}_{\text{L}}^{\prime}=\boldsymbol{H}_{\text{L}}^{00}+\boldsymbol{H}_{\text{L}}^{10}\boldsymbol{g}_{\text{L},-}^{\text{ret}}\boldsymbol{H}_{\text{L}}^{01}$
and $\boldsymbol{H}_{\text{R}}^{\prime}=\boldsymbol{H}_{\text{R}}^{00}+\boldsymbol{H}_{\text{R}}^{01}\boldsymbol{g}_{\text{R},+}^{\text{ret}}\boldsymbol{H}_{\text{R}}^{10}$
represent the left and right edge, respectively while $\boldsymbol{H}_{\text{C}}^{\prime}=\boldsymbol{H}_{\text{C}}$
and $\boldsymbol{H}_{\text{CL/CR}}^{\prime}=\boldsymbol{H}_{\text{CL/CR}}=(\boldsymbol{H}_{\text{LC/RC}}^{\prime})^{\dagger}$.
The frequency-dependent retarded surface Green's functions $\boldsymbol{g}_{\text{L},-}^{\text{ret}}$
and $\boldsymbol{g}_{\text{R},+}^{\text{ret}}$ are given by \begin{subequations}
\\
\begin{equation}
\boldsymbol{g}_{\alpha,-}^{\text{ret}}=[(\omega^{2}+i\eta)\boldsymbol{I}_{\alpha}-\boldsymbol{H}_{\alpha}^{00}-\boldsymbol{H}_{\alpha}^{10}\boldsymbol{g}_{\alpha,-}^{\text{ret}}\boldsymbol{H}_{\alpha}^{01}]^{-1}\label{eq:RetardedLeftSurfaceGF}
\end{equation}
\begin{equation}
\boldsymbol{g}_{\alpha,+}^{\text{ret}}=[(\omega^{2}+i\eta)\boldsymbol{I}_{\alpha}-\boldsymbol{H}_{\alpha}^{00}-\boldsymbol{H}_{\alpha}^{01}\boldsymbol{g}_{\alpha,+}^{\text{ret}}\boldsymbol{H}_{\alpha}^{10}]^{-1}\label{eq:RetardedRightSurfaceGF}
\end{equation}
\label{eq:AllRetardedSurfaceGF}\end{subequations} where $\eta$
is the small infinitesimal part that we add to $\omega^{2}$ to satisfy
causality, and they are commonly generated using the decimation technique~\citep{FGuinea:PRB83_Effective}
or by solving the generalized eigenvalue equation.~\citep{JSWang:EPJB08_Quantum,SSadasivam:PRB17_Phonon}
Physically, Eq.~(\ref{eq:RetardedLeftSurfaceGF}) is the retarded
surface Green's function for a decoupled semi-infinite lattice extending
infinitely to the left (denoted by the `-' in the subscript of $\boldsymbol{g}_{\alpha,-}^{\text{ret}}$)
while Eq.~(\ref{eq:RetardedRightSurfaceGF}) is the corresponding
surface Green's function for a decoupled semi-infinite lattice extending
infinitely to the right (denoted by the `+' in the subscript of $\boldsymbol{g}_{\alpha,+}^{\text{ret}}$).
In addition, the advanced surface Green's functions can be obtained
from the Hermitian conjugates of Eq.~(\ref{eq:AllRetardedSurfaceGF}),
i.e. $\boldsymbol{g}_{\alpha,-}^{\text{adv}}=(\boldsymbol{g}_{\alpha,-}^{\text{ret}}){}^{\dagger}$
and $\boldsymbol{g}_{\alpha,+}^{\text{adv}}=(\boldsymbol{g}_{\alpha,+}^{\text{ret}})^{\dagger}$. 

\subsubsection{Phonon transmittance and current}

To find the phonon transmission through the interface, we compute
the corresponding Green's function for Eq.~(\ref{eq:ProjectedForceConstantMatrix}),
$\boldsymbol{G}^{\text{ret}}=[(\omega^{2}+i\eta)\mathbf{I}^{\prime}-\mathbf{H}^{\prime}]^{-1}$
where $\mathbf{I}^{\prime}$ is an identity matrix of the same size
as $\mathbf{H}^{\prime}$; the $\boldsymbol{G}^{\text{ret}}$ matrix
can be partitioned into submatrices in the same manner as $\mathbf{H}^{\prime}$,
i.e.
\begin{equation}
\boldsymbol{G}^{\text{ret}}=\left(\begin{array}{ccc}
\boldsymbol{G}_{\text{L}}^{\text{ret}} & \boldsymbol{G}_{\text{LC}}^{\text{ret}} & \boldsymbol{G}_{\text{LR}}^{\text{ret}}\\
\boldsymbol{G}_{\text{CL}}^{\text{ret}} & \boldsymbol{G}_{\text{C}}^{\text{ret}} & \boldsymbol{G}_{\text{CR}}^{\text{ret}}\\
\boldsymbol{G}_{\text{RL}}^{\text{ret}} & \boldsymbol{G}_{\text{RC}}^{\text{ret}} & \boldsymbol{G}_{\text{R}}^{\text{ret}}
\end{array}\right)\ .\label{eq:FiniteGreensFunction}
\end{equation}
In the original AGF method,\citep{WZhang:NHT07,JSWang:EPJB08_Quantum}
the \emph{phonon transmittance} through the scattering region is given
by the well-known Caroli formula:~\citep{CCaroli:JPhysC71,WZhang:NHT07,JSWang:EPJB08_Quantum}
\begin{equation}
\Xi(\omega)=\text{Tr}[\boldsymbol{\Gamma}_{\text{R}}\boldsymbol{G}_{\text{RL}}^{\text{ret}}\boldsymbol{\Gamma}_{\text{L}}(\boldsymbol{G}_{\text{RL}}^{\text{ret}})^{\dagger}]\label{eq:CaroliFormula}
\end{equation}
 where $\boldsymbol{\Gamma}_{\text{L}}=i\boldsymbol{H}_{\text{L}}^{10}(\boldsymbol{g}_{\text{L},-}^{\text{ret}}-\boldsymbol{g}_{\text{L},-}^{\text{adv}})\boldsymbol{H}_{\text{L}}^{01}$
and $\boldsymbol{\Gamma}_{\text{R}}=i\boldsymbol{H}_{\text{R}}^{01}(\boldsymbol{g}_{\text{R},+}^{\text{ret}}-\boldsymbol{g}_{\text{R},+}^{\text{adv}})\boldsymbol{H}_{\text{R}}^{10}$,
while the total phonon heat flux between the leads is given by
\begin{equation}
J(\omega)=\int_{0}^{\infty}d\omega\frac{\hbar\omega}{2\pi}[f_{\text{L}}(\omega)-f_{\text{R}}(\omega)]\Xi(\omega)\label{eq:PhononHeatFlux}
\end{equation}
where $f_{\alpha}(\omega)=[\exp(\frac{\hbar\omega}{k_{B}T_{\alpha}})-1]^{-1}$
is the Bose-Einstein occupation factor for the $\alpha$ lead at temperature
$T_{\alpha}$.

\subsection{Recent extensions of AGF method for mode-resolved transmission}

At this point, we depart from the usual AGF method to describe how
the traditional AGF formalism can be extended to find the constituent
phonons in the heat flux in Eq.~(\ref{eq:PhononHeatFlux}). From
the Green's function $\boldsymbol{G}^{\text{ret}}$ in Eq.~(\ref{eq:FiniteGreensFunction}),
we can use the traditional AGF method to compute the phonon transmittance
$\Xi(\omega)$ which is the sum of the individual phonon transmission
coefficients.~\citep{ZHuang:JHT11_Modeling,ZYOng:PRB15_Efficient}
A more explicit connection to conventional scattering theory may be
made by noting that the individual transmission coefficients can be
derived directly from the diagonal elements of the transmission matrix,~\citep{DSFisher:PRB81_Relation}
which relates the amplitude of the incoming phonon flux to that of
the outgoing forward-scattered (or transmitted) phonon flux and is
computed numerically from $\boldsymbol{G}^{\text{ret}}$.~\citep{ZYOng:PRB15_Efficient} 

\subsubsection{Bloch matrices and bulk phonon eigenmodes}

As a prerequisite to computing the transmission coefficient of individual
phonon modes at a given $\omega$, we need to determine all the individual
phonon modes which can be derived from surface Green's functions in
Eq.~(\ref{eq:AllRetardedSurfaceGF}). We motivate the following derivation
by first pointing out that the surface Green's functions for the $\alpha$
lead depend only on $\boldsymbol{H}_{\alpha}^{00}$ and $\boldsymbol{H}_{\alpha}^{01}$,
which are also the matrices used to find the bulk phonon eigenmodes
in Eq.~(\ref{eq:BulkPhononDispersion}), suggesting that properties
of the bulk phonons are encoded in the surface Green's function matrices. 

The link between the surface Green's function and the bulk phonon
eigenmodes may be more firmly established by noting that the advanced
and retarded Bloch matrices~\citep{TAndo:PRB91_Quantum,PAKhomyakov:PRB05_Conductance,ZYOng:PRB15_Efficient}
of the left and right lead, $\boldsymbol{F}_{\alpha}^{\text{adv/ret}}(+)$
and $\boldsymbol{F}_{\alpha}^{\text{adv/ret}}(-)$, which describe
the bulk translational symmetry along the direction of the heat flux,
can be computed directly from the formulae: \begin{subequations}
\begin{equation}
\boldsymbol{F}_{\alpha}^{\text{adv/ret}}(+)=\boldsymbol{g}_{\alpha,+}^{\text{adv/ret}}\boldsymbol{H}_{\alpha}^{10}\label{eq:RightGoingBlochMatrix}
\end{equation}
\begin{equation}
\boldsymbol{F}_{\alpha}^{\text{adv/ret}}(-)^{-1}=\boldsymbol{g}_{\alpha,-}^{\text{adv/ret}}\boldsymbol{H}_{\alpha}^{01}\ .\label{eq:LeftGoingBlochMatrix}
\end{equation}
\label{eq:BlochMatrices}\end{subequations} As pointed out in Ref.~\citep{ZYOng:PRB15_Efficient},
the bulk eigenmodes for the lead at frequency $\omega$ can be determined
directly from the Bloch matrices by solving the eigenvalue equations:
\begin{subequations} 
\begin{equation}
\boldsymbol{F}_{\alpha}^{\text{adv/ret}}(+)\boldsymbol{U}_{\alpha}^{\text{adv/ret}}(+)=\boldsymbol{U}_{\alpha}^{\text{adv/ret}}(+)\boldsymbol{\Lambda}_{\alpha}^{\text{adv/ret}}(+)\label{eq:RightGoingModes}
\end{equation}
\begin{equation}
\boldsymbol{F}_{\alpha}^{\text{adv/ret}}(-)^{-1}\boldsymbol{U}_{\alpha}^{\text{adv/ret}}(-)=\boldsymbol{U}_{\alpha}^{\text{adv/ret}}(-)\boldsymbol{\Lambda}_{\alpha}^{\text{adv/ret}}(-)^{-1}\label{eq:LeftGoingModes}
\end{equation}
\label{eq:BlochMatrixEigenmodes}\end{subequations} where $\boldsymbol{U}_{\alpha}^{\text{ret}}(+)$
{[}$\boldsymbol{U}_{\alpha}^{\text{ret}}(-)${]} is a matrix with
its column vectors corresponding to the \emph{retarded} rightward-going
(leftward-going) extended or rightward (leftward) decaying evanescent
modes at frequency $\omega$ and has the form $\boldsymbol{U}_{\alpha}^{\text{ret}}=(\boldsymbol{e}_{1}\boldsymbol{e}_{2}\ldots\boldsymbol{e}_{N})$
where $\boldsymbol{e}_{n}$ is a normalized right eigenvector of the
Bloch matrix in the $n$-th column of $\boldsymbol{U}_{\alpha}^{\text{ret}}$.
Similarly, $\boldsymbol{U}_{\alpha}^{\text{adv}}(-)$ {[}$\boldsymbol{U}_{\alpha}^{\text{adv}}(+)${]}
is a matrix with its column vectors corresponding to the \emph{advanced}
rightward-going (leftward-going) extended or leftward (rightward)
decaying evanescent modes. The matrix $\boldsymbol{\Lambda}_{\alpha}^{\text{adv/ret}}(+)$
{[}$\boldsymbol{\Lambda}_{\alpha}^{\text{adv/ret}}(-)${]} is a diagonal
matrix with matrix elements of the form $e^{ik_{n}a}$ where $k_{n}$
is the phonon wave vector corresponding to the $n$-th column eigenvector
in $\boldsymbol{U}_{\alpha}^{\text{adv/ret}}(+)$ {[}$\boldsymbol{U}_{\alpha}^{\text{adv/ret}}(-)${]}
and satisfying $-\frac{\pi}{a_{\alpha}}<k_{n}\leq\frac{\pi}{a_{\alpha}}$.
At this juncture, it would appear that we have two redundant sets
of eigenmodes, $\boldsymbol{U}_{\alpha}^{\text{ret}}(\pm)$ and $\boldsymbol{U}_{\alpha}^{\text{adv}}(\pm)$,
although it can be shown later that the former (retarded modes) corresponds
to outgoing transmitted phonons while the latter (advanced modes)
corresponds to phonons that are incident on the interface.

We note that because the Bloch matrices are not Hermitian, their eigenvectors
are not necessarily orthogonal and this can be problematic for transmission
coefficient calculations~\citep{SSadasivam:PRB17_Phonon} when the
eigenvectors have the same $k$ and are wave vector-degenerate, i.e.
they have the same $\omega$ and $k$. We resolve this by orthonormalizing
the wave vector-degenerate column eigenvectors in $\boldsymbol{U}_{\alpha}^{\text{adv/ret}}$
with a Gram-Schmidt procedure.~\citep{GArfken:Book95_Mathematical,CMWerneth:EJP10_Numerical}
The final piece of ingredient needed for the calculation of the individual
phonon transmission coefficients is the diagonal eigenvelocity matrix~\citep{PAKhomyakov:PRB05_Conductance,JSWang:EPJB08_Quantum}
\begin{align}
\boldsymbol{V}_{\alpha}^{\text{adv/ret}}(+)=\frac{ia_{\alpha}}{2\omega}[\boldsymbol{U}_{\alpha}^{\text{adv/ret}}(+)]^{\dagger}\boldsymbol{H}_{\alpha}^{01}[\boldsymbol{g}_{\alpha,+}^{\text{adv/ret}}-(\boldsymbol{g}_{\alpha,+}^{\text{ret/adv}})^{\dagger}]\boldsymbol{H}_{\alpha}^{10}\boldsymbol{U}_{\alpha}^{\text{adv/ret}}(+)\ ,\label{eq:RightGoingVelocityMatrix}
\end{align}
which has the group velocities of the eigenvectors in $\boldsymbol{U}_{\alpha}^{\text{adv/ret}}(+)$
as its diagonal elements. Likewise, $\boldsymbol{V}_{\alpha}^{\text{adv/ret}}(-)$
is defined as
\begin{align}
\boldsymbol{V}_{\alpha}^{\text{adv/ret}}(-)=-\frac{ia_{\alpha}}{2\omega}[\boldsymbol{U}_{\alpha}^{\text{adv/ret}}(-)]^{\dagger}\boldsymbol{H}_{\alpha}^{10}[\boldsymbol{g}_{\alpha,-}^{\text{adv/ret}}-(\boldsymbol{g}_{\alpha,-}^{\text{ret/adv}})^{\dagger}]\boldsymbol{H}_{\alpha}^{01}\boldsymbol{U}_{\alpha}^{\text{adv/ret}}(-)\ .\label{eq:LeftGoingVelocityMatrix}
\end{align}
For the evanescent modes, the group velocity is always zero while
for propagating modes that contribute to the heat flux, the group
velocity is positive (negative) in $\boldsymbol{V}_{\alpha}^{\text{ret}}(+)$
and $\boldsymbol{V}_{\alpha}^{\text{adv}}(-)$ {[}$\boldsymbol{V}_{\alpha}^{\text{ret}}(-)$
and $\boldsymbol{V}_{\alpha}^{\text{adv}}(+)${]}. In addition, we
define the diagonal matrices $\widetilde{\boldsymbol{V}}_{\alpha}^{\text{adv/ret}}(+)$
and $\widetilde{\boldsymbol{V}}_{\alpha}^{\text{adv/ret}}(-)$ in
which their nonzero diagonal matrix elements are the inverse of those
of $\boldsymbol{V}_{\alpha}^{\text{adv/ret}}(+)$ and $\boldsymbol{V}_{\alpha}^{\text{adv/ret}}(-),$
respectively. For each lead, we can also define the diagonal matrices
\begin{subequations}
\begin{equation}
\boldsymbol{I}_{\alpha}^{\text{adv/ret}}(+)=\boldsymbol{V}_{\alpha}^{\text{adv/ret}}(+)\widetilde{\boldsymbol{V}}_{\alpha}^{\text{adv/ret}}(+)\label{eq:RightGoingIdentityMatrix}
\end{equation}
\begin{equation}
\boldsymbol{I}_{\alpha}^{\text{adv/ret}}(-)=\boldsymbol{V}_{\alpha}^{\text{adv/ret}}(-)\widetilde{\boldsymbol{V}}_{\alpha}^{\text{adv/ret}}(-)\label{eq:LeftGoingIdentityMatrix}
\end{equation}
\label{eq:AllGoingIdentityMatrices}\end{subequations} in which the
$n$-th diagonal element equals $1$ if the $n$-th column of $\boldsymbol{U}_{\alpha}^{\text{adv/ret}}(+)$
and $\boldsymbol{U}_{\alpha}^{\text{adv/ret}}(-)$ corresponds to
an extended mode and $0$ otherwise. Therefore, it follows from Eq.~(\ref{eq:AllGoingIdentityMatrices}
) that the number of rightward-going phonon channels $N_{\alpha}(+)$
and the number of leftward-going phonon channels $N_{\alpha}(-)$
are given by \begin{subequations}
\begin{equation}
N_{\alpha}(+)=\text{Tr}[\boldsymbol{I}_{\alpha}^{\text{ret}}(+)]=\text{Tr}[\boldsymbol{I}_{\alpha}^{\text{adv}}(-)]\label{eq:NumberRightGoingChannels}
\end{equation}
\begin{equation}
N_{\alpha}(-)=\text{Tr}[\boldsymbol{I}_{\alpha}^{\text{ret}}(-)]=\text{Tr}[\boldsymbol{I}_{\alpha}^{\text{adv}}(+)]\ .\label{eq:NumberLeftGoingChannels}
\end{equation}
\label{eq:AllNumberOfChannels}\end{subequations} 

\subsubsection{Phonon transmission matrices and transmission coefficients}

Now, let us consider the scattering problem for an incoming phonon
at frequency $\omega$ from the left lead that is incident on the
scattering region. In the $n=0$ slice at the edge of the left lead,
the motion of the degrees of freedom, $\boldsymbol{c}_{0}=(c_{0,1},\ldots,c_{0,N})^{T}$
where $c_{0,l}$ is the complex coefficient for the $l$th degree
of freedom in the slice for $l=1,\ldots,N$, can be decomposed into
two parts, i.e. 
\begin{equation}
\boldsymbol{c}_{0}=\boldsymbol{c}_{0}(+)+\boldsymbol{c}_{0}(-)\label{eq:Slice0Motion}
\end{equation}
where $\boldsymbol{c}_{0}(+)$ and $\boldsymbol{c}_{0}(-)$ respectively
represent the rightward-going (incident) and leftward-going (reflected)
components, while in the $n=2$ slice at the edge of the right lead,
the motion of its degrees of freedom is given by 
\begin{equation}
\boldsymbol{c}_{2}=\boldsymbol{c}_{2}(+)\ ,\label{eq:TransmittedWave}
\end{equation}
where the RHS represents a rightward-going (transmitted) $\omega$-frequency
wave which can be a linear combination of bulk right-lead phonon modes
propagating away from the interface. Suppose the rightward-going component
in Eq.~(\ref{eq:Slice0Motion}) is a left-lead bulk phonon mode,
i.e. $\boldsymbol{c}_{0}(+)=\boldsymbol{u}_{\text{L},n}(k,\omega)$
where $n$ and $k$ are the phonon polarization index and wave vector,
respectively. Then, it can be shown~\citep{PAKhomyakov:PRB05_Conductance}
that the transmitted wave $\boldsymbol{c}_{2}(+)$ in the right lead
is related to the incident wave $\boldsymbol{c}_{0}(+)$ from the
right lead, via the expression 
\begin{equation}
\boldsymbol{c}_{2}=\boldsymbol{G}_{\text{RL}}^{\text{ret}}\boldsymbol{Q}_{\text{L}}\boldsymbol{u}_{\text{L},n}(k,\omega)\label{eq:TransmittedIncidentWave}
\end{equation}
where 
\begin{align}
\boldsymbol{Q}_{\alpha}=(\omega^{2}+i\eta)\boldsymbol{I}_{\alpha}-\boldsymbol{H}_{\alpha}^{00}-\boldsymbol{H}_{\alpha}^{10}\boldsymbol{g}_{\alpha,-}^{\text{ret}}(\omega)\boldsymbol{H}_{\alpha}^{01}-\boldsymbol{H}_{\alpha}^{01}\boldsymbol{g}_{\alpha,+}^{\text{ret}}(\omega)\boldsymbol{H}_{\alpha}^{10}\label{eq:InverseBulkGreensFunction}
\end{align}
and $\boldsymbol{Q}_{\alpha}^{-1}$ is the bulk Green's function of
the $\alpha$ lead. The expression in Eq.~(\ref{eq:TransmittedIncidentWave})
can be expressed as a linear combination of transmitted right-lead
phonon modes $\boldsymbol{u}_{\text{R},m}(k_{m},\omega)$, i.e. $\boldsymbol{c}_{2}=\sum_{m}\boldsymbol{u}_{\text{R},m}(k_{m},\omega)\tau_{mn}$,
where $\tau_{mn}$ is the linear coefficient and forms the matrix
elements of the transmission matrix $\boldsymbol{\tau}$, where
\begin{equation}
\boldsymbol{\tau}=[\boldsymbol{U}_{\text{R}}^{\text{ret}}(+)]^{-1}\boldsymbol{G}_{\text{RL}}^{\text{ret}}\boldsymbol{Q}_{\text{L}}\boldsymbol{U}_{\text{L}}^{\text{ret}}(+)\ .\label{eq:taumatrix}
\end{equation}
The flux-normalized transmission matrix is $\boldsymbol{t}_{\text{RL}}=[\boldsymbol{V}_{\text{R}}^{\text{ret}}(+)]^{\nicefrac{1}{2}}\boldsymbol{\tau}[\widetilde{\boldsymbol{V}}_{\text{L}}^{\text{adv}}(-)]^{\nicefrac{1}{2}}$,
which we can rewrite as~\citep{ZYOng:PRB15_Efficient} 
\begin{align}
\boldsymbol{t}_{\text{RL}}=\frac{2i\omega}{\sqrt{a_{\text{R}}a_{\text{L}}}}[\boldsymbol{V}_{\text{R}}^{\text{ret}}(+)]^{\nicefrac{1}{2}}[\boldsymbol{U}_{\text{R}}^{\text{ret}}(+)]^{-1}\boldsymbol{G}_{\text{RL}}^{\text{ret}}[\boldsymbol{U}_{\text{L}}^{\text{adv}}(-)^{\dagger}]^{-1}[\boldsymbol{V}_{\text{L}}^{\text{adv}}(-)]^{\nicefrac{1}{2}}\ .\label{eq:tmatrix_RL}
\end{align}
Each row of $\boldsymbol{t}_{\text{RL}}$ corresponds to either a
transmitted right-lead extended or evanescent mode. For an outgoing
evanescent mode, the row elements and group velocity, given by the
corresponding diagonal element of $\boldsymbol{V}_{\text{R}}^{\text{ret}}(+)$,
are zero. Conversely, each column of of $\boldsymbol{t}_{\text{RL}}$
corresponds to either an incident left-lead extended or evanescent
mode, and the column elements and group velocity of the evanescent
modes, given by the diagonal element of $\boldsymbol{V}_{\text{L}}^{\text{adv}}(-)$.
If the $m$-th row and $n$-th column of $\boldsymbol{t}_{\text{RL}}$
correspond to extended transmitted and incident modes, then $|[\boldsymbol{t}_{\text{RL}}]_{mn}|^{2}$
gives us the probability that the incident left-lead phonon is transmitted
across the interface into the right-lead phonon. Similarly, we can
define the flux-normalized transmission matrix for phonon transmission
from the right to the left lead:

\begin{align}
\boldsymbol{t}_{\text{LR}}=\frac{2i\omega}{\sqrt{a_{\text{L}}a_{\text{R}}}}[\boldsymbol{V}_{\text{L}}^{\text{ret}}(-)]^{\nicefrac{1}{2}}[\boldsymbol{U}_{\text{L}}^{\text{ret}}(-)]^{-1}\boldsymbol{G}_{\text{LR}}^{\text{ret}}[\boldsymbol{U}_{\text{R}}^{\text{adv}}(+)^{\dagger}]^{-1}[\boldsymbol{V}_{\text{R}}^{\text{adv}}(+)]^{\nicefrac{1}{2}}\ .\label{eq:tmatrix_LR}
\end{align}

Given Eq.~(\ref{eq:tmatrix_RL}), we can construct the smaller rationalized
matrices $\bar{\boldsymbol{t}}_{\text{RL}}$ and $\bar{\boldsymbol{t}}_{\text{LR}}$
from $\boldsymbol{t}_{\text{RL}}$ and $\boldsymbol{t}_{\text{LR}}$
by deleting the matrix rows and columns corresponding to evanescent
states. This is done numerically by inspecting each diagonal element
of $\boldsymbol{I}_{\alpha}^{\text{adv/ret}}(\pm)$ of Eq.~(\ref{eq:AllGoingIdentityMatrices}),
which is either equal to 0 (evanescent) or 1 (extended), and removing
the corresponding columns or rows when $[\boldsymbol{I}_{\alpha}^{\text{adv/ret}}(\pm)]_{nn}=0$.
For example, to find $\bar{\boldsymbol{t}}_{\text{RL}}$, we inspect
$\boldsymbol{I}_{\text{R}}^{\text{ret}}(+)$ for row deletion and
$\boldsymbol{I}_{\text{L}}^{\text{adv}}(-)$ for column deletion in
$\boldsymbol{t}_{\text{RL}}$. Hence, $\bar{\boldsymbol{t}}_{\text{RL}}$
is a $N_{\text{R}}(+)\times N_{\text{L}}(+)$ matrix. Similarly, we
can also define the rationalized smaller matrices $\bar{\boldsymbol{\Lambda}}_{\alpha}^{\text{adv/ret}}(+)$
by deleting the rows and columns associated with evanescent modes
from $\boldsymbol{\Lambda}_{\alpha}^{\text{adv/ret}}(\pm)$ in Eq.~(\ref{eq:BlochMatrixEigenmodes}). 

The\emph{ transmission coefficient} of the $n$-th \emph{incoming}
phonon channel in the left lead is defined as the $n$-th diagonal
element of $\bar{\boldsymbol{t}}_{\text{RL}}^{\dagger}\bar{\boldsymbol{t}}_{\text{RL}}$,
i.e.
\begin{equation}
\Xi_{\text{L},n}=[\bar{\boldsymbol{t}}_{\text{RL}}^{\dagger}\bar{\boldsymbol{t}}_{\text{RL}}]_{nn}\ ,\label{eq:IncomingTransmitCoeff}
\end{equation}
which is equal to the fraction of its energy flux transmitted across
the interface, and its wave vector $k_{n}$ can be determined from
$[\bar{\boldsymbol{\Lambda}}_{\text{L}}^{\text{adv}}(-)]_{nn}=e^{ik_{n}a_{\text{L}}}$
or 
\begin{equation}
k_{n}=\frac{1}{a_{\text{L}}}\cos^{-1}\text{Re}[\bar{\boldsymbol{\Lambda}}_{\text{L}}^{\text{adv}}(-)]_{nn}\ .\label{eq:WaveVector}
\end{equation}
The \emph{absorption coefficient} of the $l$-th outgoing rightward-going
mode in the right lead is given by the $l$-th diagonal element of
$\bar{\boldsymbol{t}}_{\text{RL}}\bar{\boldsymbol{t}}_{\text{RL}}^{\dagger}$,
i.e.
\begin{equation}
\xi_{\text{R},l}=[\bar{\boldsymbol{t}}_{\text{RL}}\bar{\boldsymbol{t}}_{\text{RL}}^{\dagger}]_{ll}\ ,\label{eq:OutgoingTransmitCoeff}
\end{equation}
with its phonon wave vector $k_{l}$ given by $k_{l}=\frac{1}{a_{\text{R}}}\cos^{-1}\text{Re}[\bar{\boldsymbol{\Lambda}}_{\text{R}}^{\text{ret}}(+)]_{ll}$. 

The transmission coefficient for the $n$-th incoming phonon channel
in the right lead ($\Xi_{\text{R},n}=[\bar{\boldsymbol{t}}_{\text{LR}}^{\dagger}\bar{\boldsymbol{t}}_{\text{LR}}]_{nn}$)
and the absorption coefficient of the $l$-th outgoing phonon channel
in the left lead ($\xi_{\text{L},l}=[\bar{\boldsymbol{t}}_{\text{LR}}\bar{\boldsymbol{t}}_{\text{LR}}^{\dagger}]_{ll}$)
can be similarly defined like in Eqs.~(\ref{eq:IncomingTransmitCoeff})
to (\ref{eq:OutgoingTransmitCoeff}). We remark that the phonon transmittance
$\Xi(\omega)$ in Eq.~(\ref{eq:CaroliFormula}) is equal to the sum
of the transmission {[}Eq.~(\ref{eq:IncomingTransmission}){]} or
absorption {[}Eq.~(\ref{eq:OutgoingTransmission}){]} coefficients
of either lead, i.e. \begin{subequations}
\begin{align}
\Xi(\omega) & =\sum_{n=1}^{N_{\text{L}}(+)}\Xi_{\text{L},n}=\sum_{m=1}^{N_{\text{R}}(-)}\Xi_{\text{R},m}\ \label{eq:IncomingTransmission}\\
 & =\sum_{n=1}^{N_{\text{L}}(-)}\xi_{\text{L},n}=\sum_{m=1}^{N_{\text{R}}(+)}\xi_{\text{R},m}\ ,\label{eq:OutgoingTransmission}
\end{align}
\label{eq:AllPhononTransmittace}\end{subequations} as a consequence
of the conservation of probability current. 

\section{Example 1: linear atomic chain}

\begin{figure}
\includegraphics[scale=0.5]{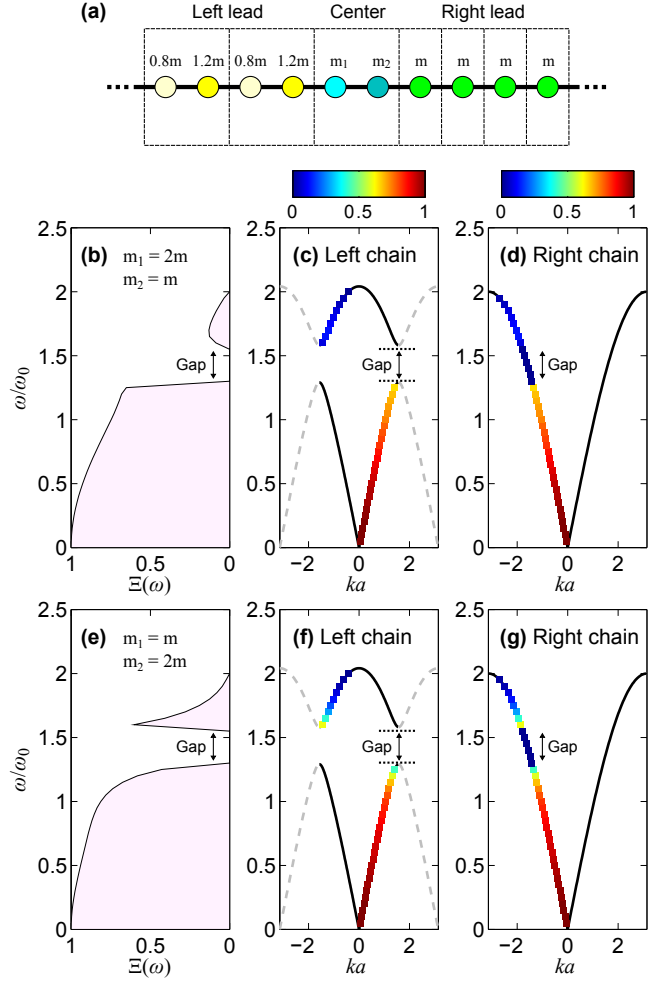}

\caption{\textbf{(a)} Schematic of junction between two linear atomic chains.
\textbf{(b)} Plot of the transmittance $\Xi(\omega)$ as a function
of frequency $\omega$ for $m_{1}=2m$ and $m_{2}=m$. The corresponding
transmission coefficients for the bulk phonon modes are superimposed
on the phonon dispersion curves for the \textbf{(c)} left and \textbf{(d)}
right chain. \textbf{(e)} Plot of the transmittance $\Xi(\omega)$
for $m_{1}=m$ and $m_{2}=2m$. The corresponding phonon transmission
coefficients for the \textbf{(f)} left and \textbf{(g)} right chain
are also shown.}

\label{fig:LinearAtomicChain}
\end{figure}

To illustrate the basic ideas of the extended AGF method, we begin
with the relatively simple example of the linear atomic chain. We
choose this toy model because it has a simple phonon dispersion relation
which allows us to ignore the effects of polarization, leading to
a more straightforward analysis of the relationship between the phonon
transmittance spectrum and the individual phonon transmission coefficients.
In the following discussion, sufficient details of our calculations
are provided to encourage the reader to reproduce the results shown
in Fig.~\ref{fig:LinearAtomicChain}. 

Figure~\ref{fig:LinearAtomicChain}(a) shows the schematic for a
linear atomic chain junction, in which the interatomic spacing equals
$a$ and the interatomic spring constant between adjacent atoms is
$\kappa$; the atomic mass however depends on position. The interface
in the center comprises of two atoms with masses $m_{1}$ and $m_{2}$,
respectively. For the left lead, we have a semi-infinite atomic chain
that has a two-atom unit cell with masses $0.8m$ and $1.2m$, respectively,
where $m$ is the characteristic atomic mass scale, and a lattice
spacing of $2a$. For the right lead, we have another semi-infinite
atomic chain which has a one-atom unit cell with the atomic mass of
$m$ and lattice spacing of $a$. Given the model in Fig.~\ref{fig:LinearAtomicChain}(a),
the harmonic matrices that we use for our AGF calculations are
\begin{align*}
\boldsymbol{H}_{\text{L}}^{00} & =\boldsymbol{M}_{\text{L}}^{-1/2}\left(\begin{array}{cc}
2\kappa & -\kappa\\
-\kappa & 2\kappa
\end{array}\right)\boldsymbol{M}_{\text{L}}^{-1/2}\\
\boldsymbol{H}_{\text{L}}^{01} & =\boldsymbol{M}_{\text{L}}^{-1/2}\left(\begin{array}{cc}
0 & 0\\
-\kappa & 0
\end{array}\right)\boldsymbol{M}_{\text{L}}^{-1/2}\\
\boldsymbol{H}_{\text{CL}} & =\boldsymbol{M}_{\text{C}}^{-1/2}\left(\begin{array}{cc}
0 & -\kappa\\
0 & 0
\end{array}\right)\boldsymbol{M}_{\text{L}}^{-1/2}\\
\boldsymbol{H}_{\text{C}} & =\boldsymbol{M}_{\text{C}}^{-1/2}\left(\begin{array}{cc}
2\kappa & -\kappa\\
-\kappa & 2\kappa
\end{array}\right)\boldsymbol{M}_{\text{C}}^{-1/2}\\
\boldsymbol{H}_{\text{CR}} & =\boldsymbol{M}_{\text{C}}^{-1/2}\left(\begin{array}{c}
0\\
-\kappa
\end{array}\right)\boldsymbol{M}_{\text{R}}^{-1/2}\\
\boldsymbol{H}_{\text{R}}^{00} & =\frac{2\kappa}{m}\\
\boldsymbol{H}_{\text{R}}^{01} & =-\frac{\kappa}{m}
\end{align*}
where $\boldsymbol{M}_{\text{L}}=\left(\begin{array}{cc}
0.8m & 0\\
0 & 1.2m
\end{array}\right)$, $\boldsymbol{M}_{\text{C}}=\left(\begin{array}{cc}
m_{1} & 0\\
0 & m_{2}
\end{array}\right)$ and $\boldsymbol{M}_{\text{R}}=m$. In addition, a characteristic
frequency scale $\omega_{0}=\sqrt{\kappa/m}$ can be defined.

We use the Caroli formula in Eq.~(\ref{eq:CaroliFormula}) to compute
the frequency-dependent transmittance $\Xi(\omega)$ as per the traditional
AGF method. For the purpose of demonstrating the extended AGF method,
Eqs.~(\ref{eq:tmatrix_RL}) to (\ref{eq:IncomingTransmitCoeff})
are used to determine the individual phonon transmission coefficients
$\Xi_{\text{L},1}$ (left lead) and $\Xi_{\text{R},1}$ (right lead),
which we compare to the transmittance calculated with the Caroli formula.
Two sets of calculations are performed: one for $m_{1}=2m$ and $m_{2}=m$
and the other for $m_{1}=m$ and $m_{2}=2m$, i.e. we swap the atomic
masses in the center. 

Figure~\ref{fig:LinearAtomicChain}(b) shows the transmittance spectrum
obtained for $m_{1}=2m$ and $m_{2}=m$. Generally, $\Xi(\omega)$
decreases with $\omega$ because low-frequency phonons are more easily
transmitted through the interface. To understand the contribution
of the left-lead phonons to the interfacial heat flux, we superimpose
the individual phonon transmission coefficients $\Xi_{\text{L},1}$
on the phonon dispersion curve for the left chain, which has two branches
(acoustic and optical) given the two-atom unit cell, in Fig.~\ref{fig:LinearAtomicChain}(c).
The frequency gap between the bottom of the optical branch and the
top of the acoustic branch is due to the difference in the atomic
masses of the unit cell and corresponds to the transmittance gap in
Fig.~\ref{fig:LinearAtomicChain}(b), indicating the lack of available
phonon channels for transmission. In Fig.~\ref{fig:LinearAtomicChain}(c),
only the phonon modes with a positive group velocity ($\partial\omega/\partial k>0$)
are shown because they contribute to the left-lead phonon flux incident
on the interface. In the acoustic branch, the transmission coefficients
$\Xi_{\text{L},1}$ are significantly closer to unity and this explains
the higher transmittance in the spectral region below the gap in Fig.~\ref{fig:LinearAtomicChain}(b).
On the other hand, in the optical branch, the transmission coefficients
are closer to zero, corresponding to the diminished transmittance
in the spectral region above the gap. 

Alternatively, phonon transmission can also be analyzed in terms of
the right-lead phonons. In Fig.~\ref{fig:LinearAtomicChain}(d),
there is only one phonon branch as the linear chain in the right lead
has a one-atom unit cell, and only the phonon modes with a negative
group velocity ($\partial\omega/\partial k>0$) are shown because
they contribute to the right-lead phonon flux incident on the interface.
The transmission coefficients $\Xi_{\text{R},1}$ for the right-lead
phonons decrease from unity gradually as the frequency increases before
dropping abruptly to zero when $\omega/\omega_{0}>1.25$. This sudden
drop is due to the absence of phonon channels in the left lead to
which the right-lead phonons can be scattered. 

Figures~\ref{fig:LinearAtomicChain}(e) to (g) show the transmittance
and transmission coefficient spectra for a different interfacial configuration
where $m_{1}=m$ and $m_{2}=2m$. The change in the values for $m_{1}$
and $m_{2}$ results in a different transmittance spectrum in Fig.~\ref{fig:LinearAtomicChain}(e).
Nevertheless, because the atomistic structure of the left and right
lead are unchanged, the phonon dispersion curves and the $\omega$-$k$
loci of the phonon modes in Figs.~\ref{fig:LinearAtomicChain}(f)
and (g) are the same as those in Figs.~\ref{fig:LinearAtomicChain}(c)
and (d). However, transmission coefficient values in Figs.~\ref{fig:LinearAtomicChain}(f)
and (g) are different because the phonon modes are scattered differently
by the center region. Figures~\ref{fig:LinearAtomicChain}(e) and
(f) show that the marked transmittance improvement in the spectral
region above the gap is due to the higher transmission coefficients
of the phonon modes near the bottom of the optical branch of the left
lead.

\section{Example 2: carbon nanotube intramolecular junction \label{sec:Example-2-CNT}}

\begin{figure}
\includegraphics[width=8cm]{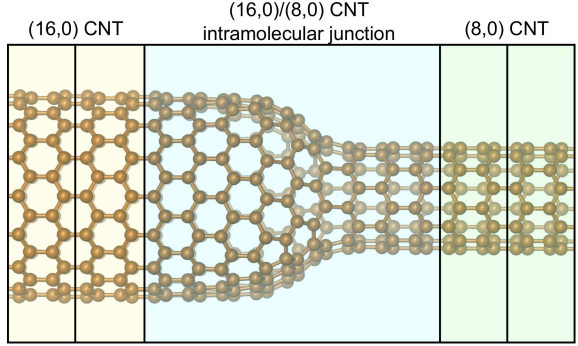}

\caption{Schematic of the left lead {[}pristine (16,0) CNT{]}, scattering region
{[}(16,0)/(8,0) CNT IMJ with 4 heptagon-pentagon defect pairs{]} and
right lead {[}pristine (8,0) CNT{]} arranged in principal layers or
slices as shown in Fig.~\ref{fig:SystemSchematic}. }
\label{fig:CNTJunction}
\end{figure}

To demonstrate the extended AGF method for a more realistic material
system, we use the technique to investigate phonon transmission and
thermal transport across the intramolecular junction (IMJ) between
a (8,0) and (16,0) carbon nanotube (CNT) as shown in Fig.~\ref{fig:CNTJunction}.
Like in Ref.~\citep{GWu:PRB07_Thermal}, two configuration of the
(16,0)/(8,0) CNT IMJ are studied, one with 4 heptagon-pentagon defect
pairs and the other with 8 heptagon-pentagon defect pairs in the IMJ.
We use the example of the CNT IMJ, which has also been studied by
Wu and Li,~\citep{GWu:PRB07_Thermal} to illustrate the level of
detail and type of insights that can be obtained from applying the
technique to the simulation of interfacial phonon transmission. 

\subsection{Generation of interatomic force-constant matrices}

The interaction between the C atoms is described by the Tersoff potential,~\citep{JTersoff:PRB89_Modeling}
with parameters taken from Ref.~\citep{LLindsay:PRB10_Optimized}.
For each configuration of the (16,0)/(8,0) CNT intramolecular junction,
three separate structures -- (1) a pristine (16,0) CNT, (2) a pristine
(8,0) CNT and (3) the (16,0)/(8,0) CNT IMJ -- are optimized using
the general utility lattice program (GULP).~\citep{JGale:MolSim03_gulp}
The interatomic force constants needed for the harmonic matrices ($\boldsymbol{H}_{\text{L}}^{00}$,
$\boldsymbol{H}_{\text{L}}^{01}$, $\boldsymbol{H}_{\text{CL}}$,
$\boldsymbol{H}_{\text{C}}$, $\boldsymbol{H}_{\text{CR}}$, $\boldsymbol{H}_{\text{R}}^{00}$
and $\boldsymbol{H}_{\text{R}}^{01}$) are computed after postprocessing
the output from GULP. We extract $\boldsymbol{H}_{\text{L}}^{00}$
and $\boldsymbol{H}_{\text{L}}^{01}$ from the pristine (16,0) CNT,
$\boldsymbol{H}_{\text{R}}^{00}$ and $\boldsymbol{H}_{\text{R}}^{01}$
from the pristine (8,0) CNT, and $\boldsymbol{H}_{\text{CL}}$, $\boldsymbol{H}_{\text{C}}$
and $\boldsymbol{H}_{\text{CR}}$ from the (16,0)/(8,0) CNT IMJ for
each IMJ configuration.

\subsection{Phonon transmittance and thermal boundary conductance}

\begin{figure}
\includegraphics[scale=0.45]{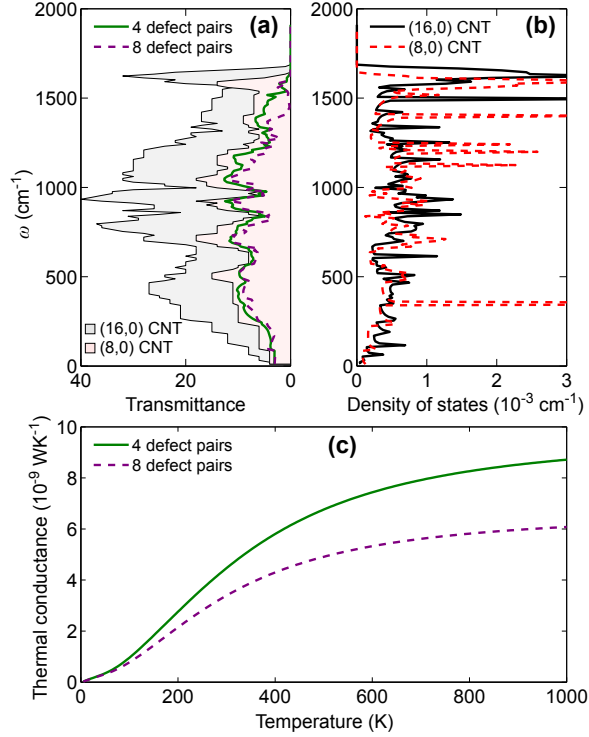}

\caption{\textbf{(a)} Phonon transmittance spectra for pristine (16,0) CNT
{[}``(16,0) CNT''{]}, pristine (8,0) CNT {[}``(8,0) CNT''{]}, the
CNT IMJ with 4 heptagon-pentagon defect pairs (``4 defect pairs''),
and the CNT IMJ with 8 defect pairs (``8 defect pairs''). \textbf{(b)}
The corresponding normalized phonon density of states for pristine
(16,0) and (8,0) CNTs. \textbf{(c)} The temperature dependence of
the thermal conductances $G_{\text{4-hp}}$ (solid line) and $G_{\text{8-hp}}$
(dashed line) from 0 to 1000 K is also shown.}

\label{fig:TraditionalAGFResults}
\end{figure}

Given the harmonic matrices ($\boldsymbol{H}_{\text{L}}^{00}$, $\boldsymbol{H}_{\text{L}}^{01}$,
$\boldsymbol{H}_{\text{CL}}$, $\boldsymbol{H}_{\text{C}}$, $\boldsymbol{H}_{\text{CR}}$,
$\boldsymbol{H}_{\text{R}}^{00}$ and $\boldsymbol{H}_{\text{R}}^{01}$),
we use Eq.~(\ref{eq:CaroliFormula}) to compute the phonon transmittance
$\Xi(\omega)$ of the (16,0)/(8,0) CNT IMJ for 4 and 8 heptagon-pentagon
defect pairs. Likewise, the total phonon transmittance for the pristine
(16,0) CNT {[}$N_{\text{L}}(+)${]} and pristine (8,0) CNT {[}$N_{\text{R}}(-)${]}
are computed from Eq.~(\ref{eq:AllNumberOfChannels}). Figure~\ref{fig:TraditionalAGFResults}(a)
shows $\Xi(\omega)$ for 4 and 8 defect pairs as well as the transmittance
$N_{\text{L}}(+)$ and $N_{\text{R}}(-)$ through the pristine (16,0)
and (18,0) CNT, respectively. In the rest of the discussion, we denote
the phonon transmittance for the CNT IMJ with $n$ defect pairs as
$\Xi_{n\text{-hp}}(\omega)$. Given that the (16,0) CNT has a larger
cross section than the (8,0) CNT, we have $N_{\text{L}}(+)>N_{\text{R}}(-)$
for all frequencies and thus, the phonon transmittance $\Xi_{n\text{-hp}}(\omega)$
through the CNT IMJ is bounded by $N_{\text{R}}(-)$ as expected.
The spectrum in Fig.~\ref{fig:TraditionalAGFResults}(a) also shows
that $\Xi_{8\text{-hp}}(\omega)<\Xi_{4\text{-hp}}(\omega)$ for almost
all $\omega$ values, consistent with Wu and Li's finding~\citep{GWu:PRB07_Thermal}
that the CNT IMJ with 8 defects pairs has a higher thermal resistance
than the CNT IMJ with 4 defects pairs. Figure~\ref{fig:TraditionalAGFResults}(b)
also shows the normalized phonon density of states for the pristine
(16,0) and (8,0) CNTs which are very similar. 

The thermal boundary conductance of the CNT IMJ with $n$ heptagon-pentagon
defect pairs can be determined using the Landauer formalism~\citep{RLandauer:JPCM89_Conductance}
and is given by 
\begin{equation}
G_{\text{IMJ}}^{n\text{-hp}}(T)=\frac{1}{2\pi}\int d\omega\ \hbar\omega\frac{df(\omega,T)}{dT}\Xi_{n\text{-hp}}(\omega)\ ,\label{Eq:ThermalBoundaryConductance}
\end{equation}
where $f(\omega,T)$ is the usual Bose-Einstein distribution function
$f(\omega,T)=[\exp(\frac{\hbar\omega}{k_{B}T})-1]^{-1}$ at temperature
$T$. We also use Eq.~(\ref{Eq:ThermalBoundaryConductance}) to compute
the temperature-dependence thermal conductance for pristine (16,0)
CNT ($G_{\text{(16,0)}}$) and pristine (8,0) CNT ($G_{\text{(8,0)}}$)
by using $N_{\text{L}}(+)$ {[}$N_{\text{R}}(-)${]} in place of $\Xi_{n\text{-hp}}(\omega)$
for $G_{\text{(16,0)}}$ ($G_{\text{(8,0)}}$). The thermal conductance
of the \emph{interface} $G_{n\text{-hp}}$ for $n$ defect pairs is
calculated using the formula~\citep{ZTian:PRB12_Enhancing,AYSerov:APL13_Effect,ZYOng:PRB15_Efficient}
\begin{equation}
G_{n\text{-hp}}=\left(\frac{1}{G_{\text{IMJ}}^{n\text{-hp}}}-\frac{1}{2G_{\text{(16,0)}}}-\frac{1}{2G_{\text{(8,0)}}}\right)^{-1}\ ,\label{Fig:InterfaceTBCFormula}
\end{equation}
assuming that the thermal resistances of the semi-infinite pristine
(16,0) CNT, the semi-infinite pristine (8,0) CNT and the (16,0)/(8,0)
CNT IMJ with $n$ defect pairs can be added in series. In the case
where both sides of the interface are of the same material, the inverse
of the RHS in Eq.~(\ref{Fig:InterfaceTBCFormula}) becomes zero as
expected, i.e. no thermal resistance is associated with the interface.
Figure~\ref{fig:TraditionalAGFResults} (c) shows the thermal conductances
$G_{\text{4-hp}}$ and $G_{\text{8-hp}}$ rising monotonically with
temperature because of the greater phonon population at higher temperatures.
In addition, we have $G_{\text{4-hp}}>G_{\text{8-hp}}$ which confirms
the findings in Ref.~\citep{GWu:PRB07_Thermal}.

\subsection{Modal dependence of phonon transmission}

Although Fig.~\ref{fig:TraditionalAGFResults}(a) yields frequency-dependent
transmittance information, we cannot discern from it the modal dependence
of individual phonon transmission. Instead, we use Eqs.~(\ref{eq:tmatrix_RL})
to (\ref{eq:IncomingTransmitCoeff}) to determine the individual phonon
transmission coefficients. At each frequency $\omega$, the transmission
coefficient and wave vector of each propagating mode in the (16,0)
CNT are determined from $\Xi_{\text{L},n}$, where $n=1,\ldots,N_{\text{L}}(+)$,
in Eq.\ (\ref{eq:IncomingTransmitCoeff}) and (\ref{eq:WaveVector}),
respectively. The phonon mode transmission coefficients $\Xi_{\text{R},m}$,
where $m=1,\ldots,N_{\text{R}}(-)$, and wave vectors in the (8,0)
CNT are similarly obtained. 

\begin{figure}
\includegraphics[scale=0.45]{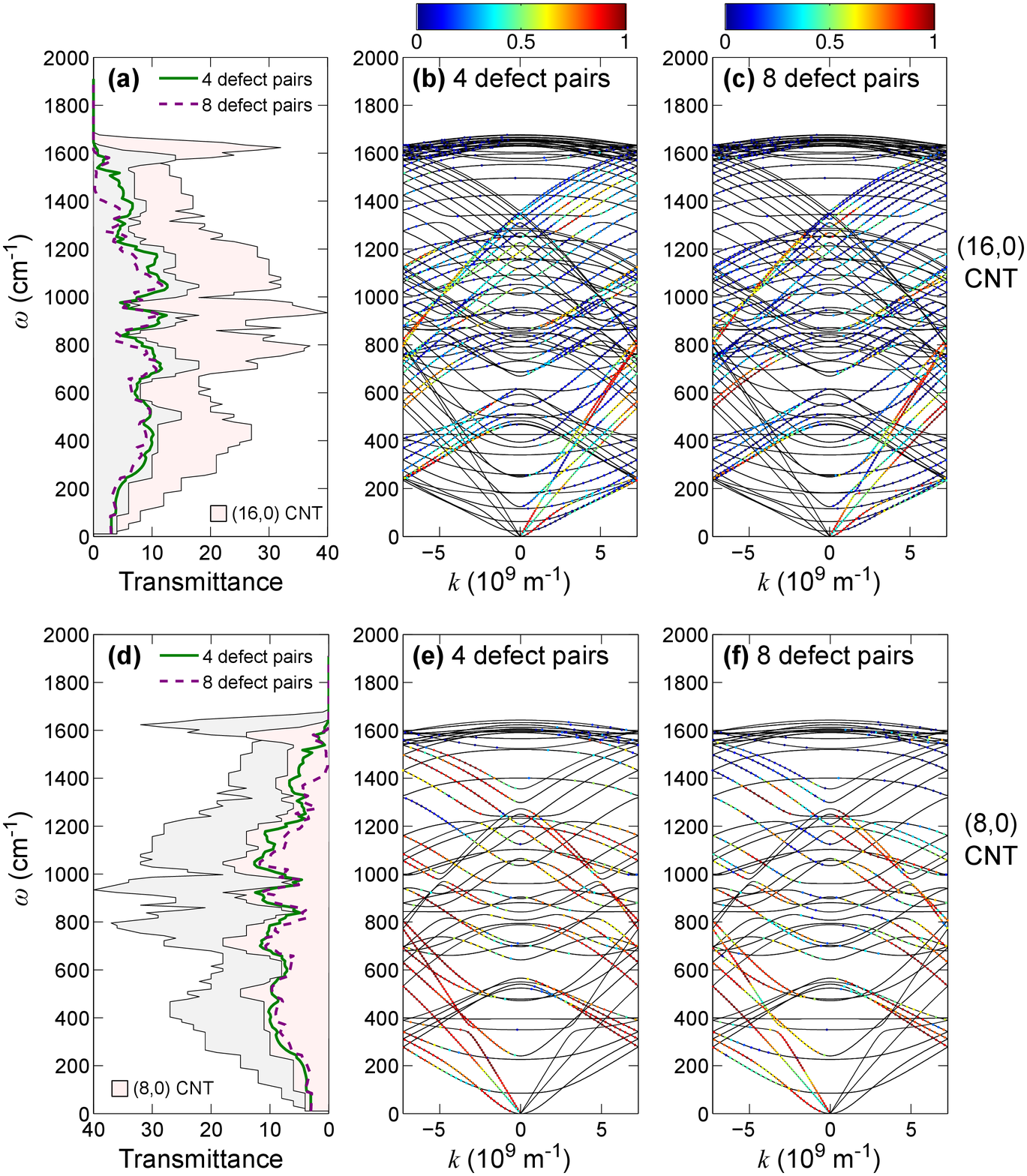}

\caption{\textbf{(a)} Plot of the left-lead phonon transmission coefficient
sum $\sum_{n=1}^{N_{\text{L}}(+)}\Xi_{\text{L},n}$ for the CNT IMJ
with 4 (green solid line) and 8 (purple dashed line) heptagon-pentagon
defect pairs, together with $N_{\text{L}}(+)$, the number of incident
channels in the (16,0) CNT (pink), and $N_{\text{R}}(-)$, the number
of channels in the (8,0) CNT (gray). The distribution of the transmission
coefficients $\Xi_{\text{L},n}$, represented in color, is superimposed
on the phonon dispersion curves of the (16,0) CNT for the CNT IMJ
with \textbf{(b)} 4 and \textbf{(c)} 8 heptagon-pentagon defect pairs.
\textbf{(b)} Plot of the right-lead phonon transmission coefficient
sum $\sum_{m=1}^{N_{\text{R}}(-)}\Xi_{\text{R},m}$ for the CNT IMJ
with 4 (green solid line) and 8 (purple dashed line) heptagon-pentagon
defect pairs, together with $N_{\text{R}}(-)$, the number of incident
channels in the (8,0) CNT (pink), and $N_{\text{L}}(+)$, the number
of channels in the (16,0) CNT (gray). The distribution of the transmission
coefficients $\Xi_{\text{R},m}$ is superimposed on the phonon dispersion
curves of the (8,0) CNT for the CNT IMJ with \textbf{(b)} 4 and \textbf{(c)}
8 heptagon-pentagon defect pairs.}

\label{fig:ModalTransmissionSpectra}
\end{figure}

Figure~\ref{fig:ModalTransmissionSpectra}(a) shows the left-lead
phonon transmission coefficient sum $\sum_{n=1}^{N_{\text{L}}(+)}\Xi_{\text{L},n}$,
describing the left-to-right phonon flux from the (16,0) CNT to the
(8,0) CNT across the CNT IMJ with 4 and 8 heptagon-pentagon defect
pairs, as well as $N_{\text{L}}(+)$ and $N_{\text{R}}(-)$ for the
pristine (16,0) and (8,0) CNT, respectively. As expected from Eq.~(\ref{eq:IncomingTransmission}),
the transmission coefficient sum $\sum_{n=1}^{N_{\text{L}}(+)}\Xi_{\text{L},n}$
in Fig.~\ref{fig:ModalTransmissionSpectra}(a) is identical to the
phonon transmittance spectrum $\Xi_{n\text{-hp}}(\omega)$ in Fig.~\ref{fig:TraditionalAGFResults}(a)
and is also larger for the CNT IMJ with 4 heptagon-pentagon defect
pairs than with 8 defect pairs. In Figs.~\ref{fig:ModalTransmissionSpectra}(b)
and (c), the numerous phonon branches associated with the phonon subband
quantization of the (16,0) CNT can be seen in the phonon dispersion
curves. The origin of the greater transmittance for 4 defect pairs
can be discerned from Figs.~\ref{fig:ModalTransmissionSpectra}(b)
and (c) which show the transmission coefficient spectra $\Xi_{\text{L},n}$
superimposed on the phonon dispersion curves of the (16,0) CNT. 

For the other side of the CNT IMJ, Fig.~\ref{fig:ModalTransmissionSpectra}(d)
shows the right-lead phonon transmission coefficient sum $\sum_{m=1}^{N_{\text{R}}(-)}\Xi_{\text{R},m}$,
describing the right-to-left phonon flux from the (8,0) CNT to the
(16,0) CNT, as well as $N_{\text{L}}(+)$ and $N_{\text{R}}(-)$ for
the pristine (16,0) and (8,0) CNT, respectively, like in Fig.~\ref{fig:ModalTransmissionSpectra}(a).
The $\sum_{m=1}^{N_{\text{R}}(-)}\Xi_{\text{R},m}$ spectrum in Fig.~\ref{fig:ModalTransmissionSpectra}(d)
is identical to the $\sum_{n=1}^{N_{\text{L}}(+)}\Xi_{\text{L},n}$
spectrum in Fig.~\ref{fig:ModalTransmissionSpectra}(a), in agreement
with Eq.~(\ref{eq:IncomingTransmission}). Hence, we can also explain
the greater phonon transmittance for the CNT IMJ with 4 defect pairs
in terms of the transmission coefficients of the (8,0) CNT phonons.
Figures~\ref{fig:ModalTransmissionSpectra}(e) and (f) show the transmission
coefficient spectra $\Xi_{\text{R},m}$ superimposed on the phonon
dispersion curves of the (8,0) CNT, which has fewer phonon branches
than the (16,0) CNT. We find from Fig.~\ref{fig:ModalTransmissionSpectra}(e)
that most of the (8,0) CNT phonons are transmitted with a near-unity
transmission coefficient. This is partly because there are fewer phonon
channels contributing to the interfacial heat flux for the (8,0) CNT
than in the (16,0) CNT and thus each (8,0) CNT phonon channel has
to transmit on average a greater percentage of its energy across the
CNT IMJ than each (16,0) CNT phonon channel. We can also tell from
comparing Figs.~\ref{fig:ModalTransmissionSpectra}(e) and (f) which
phonon branches contribute more to the interfacial phonon flux for
the CNT IMJ with 4 defect pairs than for the CNT IMJ with 8 defect
pairs, especially given the less crowded phonon dispersion curves
for the (8,0) CNT.

\begin{figure}
\includegraphics[scale=0.5]{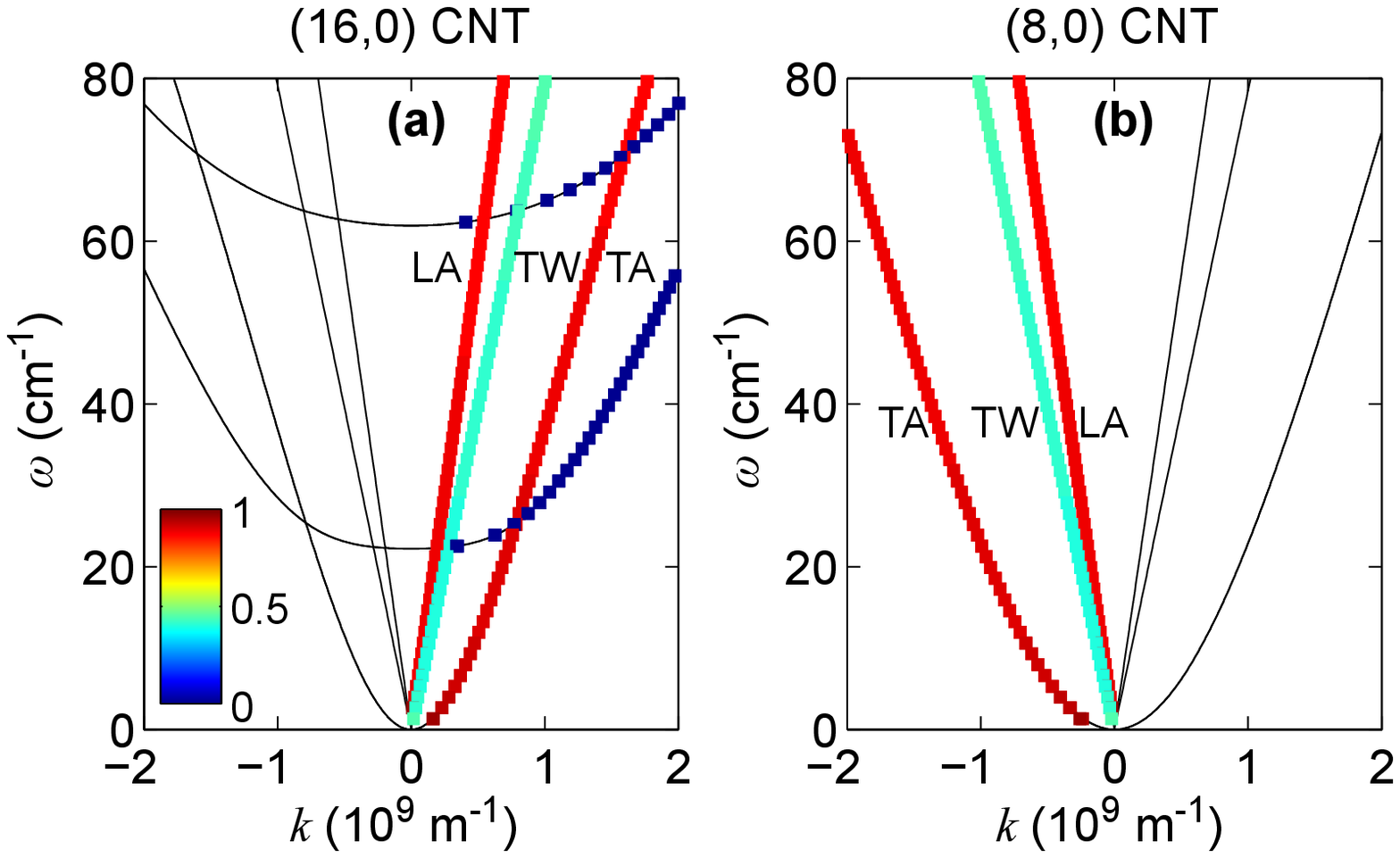}

\caption{Plot of the low-frequency phonon transmission coefficient spectra
in \textbf{(a)} (16,0) CNT and \textbf{(b)} (8,0) CNT for the CNT
IMJ with 4 heptagon-pentagon defect pairs. The longitudinal (``LA''),
torsional (``TW'') and transverse (``TA'') acoustic phonon branches
are labelled accordingly.}

\label{fig:LowFrequencySpectra}
\end{figure}

The polarization dependence of phonon transmission can also be determined
from the transmission coefficient spectra. Figure~\ref{fig:LowFrequencySpectra}
shows the low-frequency transmission coefficient spectra of the (16,0)
and (8,0) CNT for the CNT IMJ with 4 defect pairs. At very low frequencies
($\omega\rightarrow0$), there are four acoustic phonon branches in
the carbon nanotube: the longitudinal acoustic (LA), the torsional
(TW) and the doubly-degenerate transverse acoustic (TA) phonons. Our
results in Fig.~\ref{fig:LowFrequencySpectra} show that the LA and
TA phonons are transmitted across the interface with near unity transmission
probability while the TW phonons, otherwise known as ``twistons'',~\citep{PBAllen:NL07_Phonons}
are only partially transmitted with a transmission probability significantly
less than 0.5. This suggests that the CNT IMJ restricts torsional
motion even in the long-wavelength limit, resulting in the partial
reflection of TW phonons.

\section{Summary}

\begin{figure}
\includegraphics[scale=0.35]{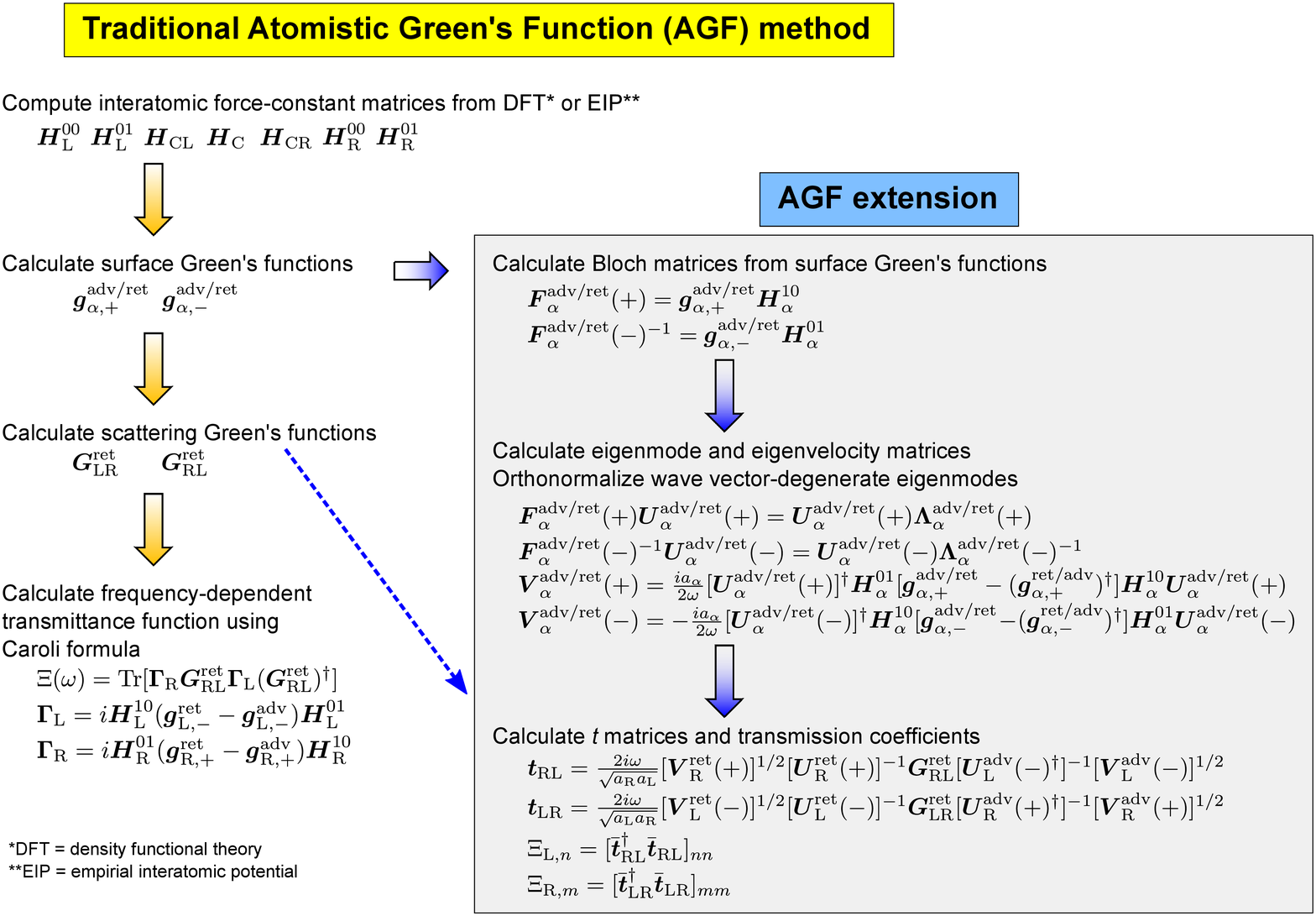}

\caption{Flow chart summarizing the main steps in the traditional AGF method
and its extension (enclosed in a gray subpanel) for computing transmission
coefficients.}

\label{fig:SummaryFlowChart}
\end{figure}

In this tutorial, we have presented an extension of the AGF method
for computing individual phonon transmission coefficients, which we
summarize in Fig.~\ref{fig:SummaryFlowChart}. In the traditional
AGF method, we use the input matrices ($\boldsymbol{H}_{\text{C}}$,
$\boldsymbol{H}_{\text{CL}}$, $\boldsymbol{H}_{\text{CR}}$, $\boldsymbol{H}_{\text{L}}^{00}$,
$\boldsymbol{H}_{\text{L}}^{01}$, $\boldsymbol{H}_{\text{R}}^{00}$
and $\boldsymbol{H}_{\text{R}}^{01}$) to compute the phonon transmittance
$\Xi(\omega)$ {[}Eq.~(\ref{eq:CaroliFormula}){]} from the surface
Green's functions $\boldsymbol{g}_{\text{L},-}^{\text{ret}}$ and
$\boldsymbol{g}_{\text{R},+}^{\text{ret}}$ {[}Eq.~(\ref{eq:AllRetardedSurfaceGF}){]}
corresponding to the decoupled leads and the retarded Green's function
of the scattering region $\boldsymbol{G}_{\text{RL}}^{\text{ret}}$
{[}Eq.~(\ref{eq:FiniteGreensFunction}){]}. In our extended AGF approach,
we exploit Eq.~(\ref{eq:BlochMatrices}) to extract the Bloch matrices
$\bm{F}_{\alpha}^{\text{adv/ret}}(\pm)$ from the surface Green's
function $\boldsymbol{g}_{\alpha,\pm}^{\text{adv/ret}}$. This allows
us to determine all the bulk phonon modes of the leads, $\bm{U}_{\alpha}^{\text{adv}}(\pm)$
and $\bm{U}_{\alpha}^{\text{ret}}(\pm)$ in Eq.~(\ref{eq:BlochMatrixEigenmodes}),
that constitute the available incoming and outgoing transmission channels
at frequency $\omega$, and their Bloch factors, $\bm{\Lambda}_{\alpha}^{\text{adv}}(\pm)$
and $\bm{\Lambda}_{\alpha}^{\text{ret}}(\pm)$. The flux-normalized
transmission matrices $\boldsymbol{t}_{\text{RL}}$ and $\boldsymbol{t}_{\text{LR}}$
which govern the transition probability between each pair of incoming
and outgoing phonon channels can be computed from Eqs.~(\ref{eq:tmatrix_RL})
and (\ref{eq:tmatrix_LR}). The single-mode transmission coefficients
$\Xi_{\text{L},n}$, which determines the phonon transmission probability,
can be computed from Eq.~(\ref{eq:IncomingTransmitCoeff}). The extended
AGF method provides a clear and detailed view of the contribution
of individual phonon modes as well as that of entire acoustic and
optical phonon branches to the thermal boundary conductance. It is
a powerful and convenient method for analyzing the effect of the atomistic
structure of the interface on the contribution of individual phonon
modes to interfacial thermal transport.

We use the simple example of a linear atomic chain junction to demonstrate
the basic ideas of the extended AGF method as well as to highlight
the connection between the phonon transmission coefficients and the
phonon transmittance as calculated with traditional AGF technique.
To illustrate the advantages of the extended AGF method for realistic
material systems, we have applied it to the study of thermal conduction
across the (16,0)/(8,0) carbon nanotube intramolecular junction. Our
analysis of phonon transmission across the CNT IMJ shows that the
transmission probability depends strongly on the CNT phonon frequency,
polarization and wave vector as well as the atomistic configuration
of the interface (i.e. the number of heptagon-pentagon defect pairs).
Our AGF simulation results suggest that phonon are more easily transmitted
across the CNT IMJ with 4 defect pairs than with 8 defect pairs, in
agreement with the findings of Ref.~\citep{GWu:PRB07_Thermal}. They
also demonstrate how the extended AGF method can play synergistic
role to other simulation-based approaches to thermal transport. 
\begin{acknowledgments}
This work was supported in part by a grant from the Science and Engineering
Research Council (Grant No. 152-70-00017) and financial support from
the Agency for Science, Technology and Research (A{*}STAR), Singapore.
I thank my colleague in the Institute of High Performance Computing
Dr. Gang Wu for providing me with the atomic coordinates of the carbon
nanotube intramolecular junctions described in Sec.~\ref{sec:Example-2-CNT}. 
\end{acknowledgments}

\appendix
\bibliographystyle{apsrev4-1}
\bibliography{PhononNotes}

\end{document}